\documentclass[conference]{IEEEtran}
\IEEEoverridecommandlockouts

\usepackage{amsmath}
\usepackage{graphicx}
\usepackage{cite}
\usepackage{amsfonts}
\usepackage{amsthm}
\usepackage{algorithm}
\usepackage{algorithmic}
\usepackage{amssymb}

\theoremstyle{plain}  
\theoremstyle{plain}  
\theoremstyle{plain}

\begin{document}

\title{Securing UAV Communications Via Trajectory Optimization\thanks{This work was supported in part by the National Natural Science Foundation of China under Grand 61571138, the Natural Science Foundation of Guangdong Province under Grand 2015A030313481, the Science and Technology Plan Project of Guangdong Province under Grands 2016A050503044, 2016KZ010107, 2016B090904001, and 2016B090918031, the Science and Technology Plan Project of Guangzhou City under Grand 201604020127, and the scholarship from China Scholarship Council under Grant 201608440436.}}

\author{\IEEEauthorblockN{Guangchi Zhang\IEEEauthorrefmark{1}\IEEEauthorrefmark{2},
Qingqing Wu\IEEEauthorrefmark{2},
Miao Cui\IEEEauthorrefmark{1},
and Rui Zhang\IEEEauthorrefmark{2}}
\IEEEauthorblockA{\IEEEauthorrefmark{1}School of Information Engineering,
Guangdong University of Technology}
\IEEEauthorblockA{\IEEEauthorrefmark{2}Department of Electrical and Computer Engineering, National University of Singapore\\
Email: gczhang@gdut.edu.cn, elewuqq@nus.edu.sg, cuimiao@gdut.edu.cn, elezhang@nus.edu.sg} }

\maketitle

\begin{abstract}
Unmanned aerial vehicle (UAV) communications has drawn significant interest recently due to many advantages such as low cost, high mobility, and on-demand deployment. This paper addresses the issue of physical-layer security in a UAV communication system, where a UAV sends confidential information to a legitimate receiver in the presence of a potential eavesdropper which are both on the ground. We aim to maximize the secrecy rate of the system by jointly optimizing the UAV's trajectory and transmit power over a finite horizon. In contrast to the existing literature on wireless security with static nodes, we exploit the mobility of the UAV in this paper to enhance the secrecy rate via a new trajectory design. Although the formulated problem is non-convex and challenging to solve, we propose an iterative algorithm to solve the problem efficiently, based on the block coordinate descent and successive convex optimization methods. Specifically, the UAV's transmit power and trajectory are each optimized with the other fixed in an alternating manner until convergence. Numerical results show that the proposed algorithm significantly improves the secrecy rate of the UAV communication system, as compared to benchmark schemes without transmit power control or trajectory optimization.
\end{abstract}

\IEEEpeerreviewmaketitle


\section{Introduction}
With many advantages such as high mobility, low cost, and on-demand deployment, unmanned aerial vehicles (UAVs) have been more widely used in both military and civilian applications such as transport, surveillance, search and rescue, etc. Recently, UAVs have also found increasingly more applications in wireless communications \cite{Zeng2016}, such as mobile coverage \cite{Mozaffari2015, Bor2016, Lyu2017, WuAPCC2017, WuGC2017}, mobile relaying \cite{Zeng2016Trans, Johansen2014}, mobile data collection \cite{Sotheara2014, Lyu2016}, etc. In UAV-based wireless communication systems, how to secure the transmission of confidential information against intentional or unintentional eavesdropping is an important problem. In this paper, we focus our study on physical-layer security, which has been extensively investigated in wireless communication as a viable anti-eavesdropping technique. The key design metric in physical-layer security that has been widely adopted is the so-called secrecy rate (e.g. \cite{Goel2008, QiangLi2015, My2016, Zheng2013, Gopala2008}), at which confidential message can be reliably transmitted without having the eavesdropper infer any information about the message. 

A non-zero secrecy rate can be achieved when the strength of the legitimate link is stronger than that of the eavesdropping link. In the existing literature on PHY layer security, communication nodes are usually assumed to be at fixed or quasi-static locations. As a result, the average channel quality of the legitimate/eavesdropping link mainly depends on the path loss and shadowing from the transmitter to receiver, which are predetermined if the locations of the legitimate transmitter/receiver and the eavesdropper are given. Thus, in the case that the average channel gain of the legitimate receiver is smaller than that of the eavesdropper (e.g., due to longer distance from the legitimate transmitter), in order to achieve positive secrecy rates, the  exploitation of the wireless channel small-scale fading in time, frequency and/or space becomes essential, and various techniques such as power control \cite{Gopala2008}, artificial noise (AN) \cite{Goel2008}, cooperative jamming \cite{Zheng2013}, as well as multi-antenna beamforming \cite{QiangLi2015, My2016} have been investigated. However, there are still two major challenges that remain unsolved in the existing literature. First, the achievable secrecy rate can be limited if the distance between the legitimate transmitter and its intended receiver is fixed and significantly larger than that between it and a potential eavesdropper, even if the techniques mentioned above are applied. Second, the channel state information (CSI) of the eavesdropper is usually required at the legitimate transmitter for the implementation of the above techniques, which is challenging since the eavesdropper is generally a passive device and it is thus difficult to estimate its CSI. 

In this paper, we study PHY layer security in UAV communications, which may potentially overcome the above two critical issues in conventional studies. First, in contrast to the existing literature with fixed or quasi-static nodes, the mobility of UAV is exploited to proactively establish stronger legitimate links with the ground nodes and/or degrade the channels of the eavesdroppers, by flying closer/farther to/from them, respectively, in its trajectory design. This is because in the UAV communications, the line-of-sight (LoS) channels of the UAV-to-ground links are usually much more dominant over other channel impairments such as shadowing or small-scale fading due to the much larger height of the UAV than the ground nodes. Furthermore, the UAV can practically obtain the channel gain to any potential eavesdropper on the ground by obtaining its location, which thus resolves the eavesdropper-CSI issue in the existing literature. Note that the location of any ground node as a potential eavesdropper can be detected and tracked by the UAV via using an optical camera or synthetic aperture radar (SAR) equipped on the UAV \cite{UAVSAR}.

\begin{figure}[!t]
	\centering
	\includegraphics[width=\columnwidth]{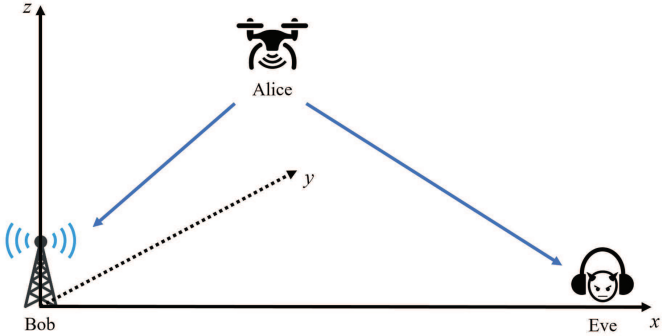}
	\caption{A UAV wireless communication system consisting of a UAV transmitter (Alice) above ground, a legitimate receiver (Bob) on the ground, and an eavesdropper (Eve) on the ground.}  \label{Figphysical}
\end{figure}

For the purpose of exposition, we consider a simplified system as shown in Fig. \ref{Figphysical}, where a UAV is deployed to send confidential information to a legitimate receiver in the presence of an eavesdropper, while the legitimate receiver and the eavesdropper are both at fixed positions on the ground. Our goal is to maximize the average secrecy rate over a finite flight period of the UAV by jointly designing the UAV trajectory and transmit power control, subject to the average and peak transmit power constraints, as well as the practical mobility constraints on the UAV's maximum speed and its initial and final locations. The formulated joint trajectory design and power control problem for secrecy rate maximization is non-convex and thus difficult to be solved optimally. To tackle this challenge, we propose an efficient algorithm by applying the block coordinate descent and successive convex optimization methods to find an approximate solution for the formulated problem. The proposed algorithm divides the design variables into two blocks, which correspond to UAV transmit power and trajectory over time, respectively, and updates the two blocks of variables one at each time with the other fixed alternately until the algorithm converges. Numerical results show that the proposed joint design can significantly improve the secrecy rate of UAV communication system, as compared to benchmark schemes without applying the trajectory optimization or transmit power control.

It is worth noting that there have been prior works (e.g. \cite{Jiang2012, Zeng2016Trans, WuTWC2017, Zeng2017}) on trajectory optimization for various UAV communication systems, which consider different system setups and performance metrics. In \cite{Zeng2016Trans}, a UAV-enabled mobile relaying system is investigated, where the UAV trajectory and transmit power are jointly designed to maximize the throughput. In \cite{Jiang2012}, the UAV flying heading is optimized to maximize the achievable sum rate from ground nodes to a UAV by assuming a constant flying speed. In \cite{WuTWC2017}, a UAV-enabled base station system serving multiple ground users is investigated, where the UAV trajectory and user scheduling are jointly optimized to maximize the minimum throughput of all users. In \cite{Zeng2017}, a new design paradigm that jointly considers both the communication throughput and the UAV's flying energy consumption is studied to maximize the energy efficiency of a point-to-point UAV-ground communication system. Different from these prior studies, in this paper we apply trajectory optimization and transmit power control to maximize the secrecy rate of a UAV wireless communication system.

\section{System Model}
As shown in Fig. \ref{Figphysical}, we consider a wireless communication system where a UAV (Alice) transmits information to a legitimate receiver (Bob) in the presence of a passive eavesdropper (Eve). We assume that Bob and Eve are both on the ground and have fixed locations, with a distance of $L$ meters (m) between them.

Without loss of generality, we consider a three-dimensional (3D) Cartesian coordinate system with Bob and Eve located at $(0,0,0)$ and $(L,0,0)$, respectively. For simplicity, we ignore the take-off and landing phases of Alice, and focus on its operation period over a finite duration $T$. It is assumed that Alice flies at a fixed altitude $H$ above ground, which can be considered as the minimum altitude required for building avoidance. The time-varying coordinate of Alice can be denoted as $( x(t),y(t),H )$, $0 \leq t \leq T$. For convenience, we divide the period $T$ into $N$ time slots with equal length, i.e., $T=N d_t$, with $d_t$ denoting the length of a time slot, which is sufficiently small such that Alice's location can be regarded as constant within each time slot. As a result, Alice's coordinate in time slot $n$ can be denoted as $( x[n],y[n],H )$. Thus, Alice's horizontal trajectory $(x(t),y(t))$ over the flight period $T$ can be approximated by the sequence $\{ x[n],y[n] \}_{n=1}^{N}$. Denote the maximum speed of Alice as $v_{\max} > 0$. The maximum flying distance of Alice in each time slot is thus $V=v_{\max} d_t$. The initial and final locations of Alice are denoted by $(x_0,y_0,H)$ and $(x_F,y_F,H)$, respectively, which are given based on the specific mission of Alice. As a result, the mobility constraints of Alice can be expressed as
\begin{subequations}  \label{EquMobilityCon}
\begin{align}
&( x[1] - x_0)^2 + (y[1] - y_0)^2 \leq V^2,  \\
&( x[n+1]-x[n] )^2 + ( y[n+1]-y[n] )^2 \leq V^2, \nonumber \\
&\quad \quad  \quad \quad n=1,\ldots,N-1,  \\
&( x_F - x[N] )^2 + (y_F - y[N])^2 \leq V^2.
\end{align}
\end{subequations}

We assume the channel from Alice to Bob and that from Alice to Eve are dominated by LoS links. Thus the channel power gain from Alice to Bob at time slot $n$ follows the free-space path loss model given by
\begin{equation}
g_{\text{AB}}[n] = \beta_0 d_{\text{AB}}^{-2}[n] = \frac{ \beta_0 }{ x^2[n] + y^2[n] + H^2 },
\end{equation}
where $\beta_0$ denotes the channel power gain at the reference distance $d_0=1$m, which depends on the carrier frequency and the antenna gains of the transmitter and receiver, and $d_{\text{AB}}[n] = \sqrt{ x^2[n] + y^2[n] + H^2 }$ is the distance from Alice to Bob at time slot $n$. Denote $P[n]$ as the transmit power of Alice at time slot $n$, which is constrained by both average and peak power limits, denoted by $\bar{P}$ and $P_{\text{peak}}$, respectively. The transmit power constraints of Alice can be expressed as
\begin{subequations} \label{EquPowerCon}
\begin{align}
&\frac{1}{N} \sum_{n=1}^{N} P[n] \leq \bar{P}, \\
&0 \leq P[n] \leq P_{\text{peak}}, \; \forall n.
\end{align}
\end{subequations}
To make the constraints above non-trivial, we assume $\bar{P}<P_{\text{peak}}$ in this paper. Without the presence of Eve, the achievable rate from Alice to Bob in bits/sec/Hertz (bps/Hz) at time slot $n$ can be expressed as
\begin{equation}  \label{EquRAB}
\begin{split}
R_{\text{AB}}[n] = & \log_2 \left(1 + \frac{ P[n] g_{\text{AB}}[n] }{ \sigma^2 }  \right)   \\
= &   \log_2 \left( 1+ \frac{ \gamma_0 P[n] } {  x^2[n] + y^2[n] + H^2 }  \right),
\end{split}
\end{equation}
where $\sigma^2$ is the additive white Gaussian noise (AWGN) power at the Bob receiver and $\gamma_0 =\beta_0 / \sigma^2$ is the reference signal-to-noise ratio (SNR). Similarly, the achievable rate from Alice to Eve in bps/Hz at time slot $n$ is given by
\begin{equation}  \label{EquRAE}
R_{\text{AE}}[n] = \log_2 \left( 1+ \frac{ \gamma_0 P[n] }{ (x[n]-L)^2 + y^2[n] + H^2 } \right).
\end{equation}
With \eqref{EquRAB} and \eqref{EquRAE}, the average secrecy rate from Alice to Bob in bps/Hz over $N$ time slots is given by \cite{Gopala2008}
\begin{equation}   \label{EquSecrecyRate}
R_{\text{sec}} = \frac{1}{N} \sum_{n=1}^N \left[ R_{\text{AB}}[n] - R_{\text{AE}}[n] \right]^{+},
\end{equation}
where $[x]^+ \triangleq \max(x,0)$.

Our objective is to maximize the average secrecy rate $R_{\text{sec}}$ given in \eqref{EquSecrecyRate} subject to Alice's mobility constraints in \eqref{EquMobilityCon} and the average and peak power constraints in \eqref{EquPowerCon}. The optimization variables include the transmit power over slots $\mathbf{P} \triangleq \left[P[1],\ldots,P[N] \right]^{\dagger}$ and Alice's trajectory in terms of its $x$ and $y$ axis position coordinates in different slots $\mathbf{x} \triangleq \left[x[1], \ldots, x[N]\right]^{\dagger}$ and $\mathbf{y} \triangleq \left[y[1], \ldots, y[N]\right]^{\dagger}$, respectively, where the superscript $\dagger$ denotes the transpose operation. Thus, we formulate the secrecy rate maximization problem as (by ignoring the constant term $1/N$ in \eqref{EquSecrecyRate})
\begin{align}
\text{(P1)}:  & \max_{ \mathbf{x}, \mathbf{y}, \mathbf{P} } \; \sum_{n=1}^N \bigg[ \log_2 \left( 1+ \frac{ \gamma_0 P[n] } {  x^2[n] + y^2[n] + H^2 }  \right) \nonumber \\
& \quad \quad - \log_2 \left(1+ \frac{ \gamma_0 P[n] } { (x[n]-L)^2 + y^2[n] + H^2 }  \right) \bigg]^+  \label{EquOriginal} \\
& \; \; \; \text{s.t.} \; \;  \eqref{EquMobilityCon}, \eqref{EquPowerCon}. \nonumber
\end{align}
Problem (P1) is difficult to be solved optimally due to the following two reasons. First, the operator $[\cdot]^+$ makes the objective function non-smooth at point zero. Second, even without $[\cdot]^+$, the objective function is still non-concave with respect to either $\mathbf{x}$, $\mathbf{y}$, or $\mathbf{P}$ in general. As a result, it is challenging to find the optimal solution to (P1) efficiently. Thus, we propose an iterative algorithm to solve it in the next section.

\section{Proposed Algorithm for Problem (P1)}
It can be shown that problem (P1) can be transformed into the following problem equivalently, i.e.,
\begin{align}
\text{(P2)}:  & \max_{ \mathbf{x}, \mathbf{y}, \mathbf{P} } \; \sum_{n=1}^N \log_2 \left( 1+ \frac{ \gamma_0 P[n] } {  x^2[n] + y^2[n] + H^2 }  \right) \nonumber \\
& \quad \quad - \log_2 \left(1+ \frac{ \gamma_0 P[n] } { (x[n]-L)^2 + y^2[n] + H^2 }  \right)   \\
&\; \; \; \text{s.t.} \; \;  \eqref{EquMobilityCon}, \eqref{EquPowerCon}. \nonumber
\end{align}
Problems (P1) and (P2) have the same optimal solution due to the fact that the value of each summation term in the objective function of both problems, i.e., $R_{\text{AB}}[n] - R_{\text{AE}}[n]$, $\forall n$, must be non-negative at optimality. This is because if $R_{\text{AB}}[n] - R_{\text{AE}}[n]<0$ for any $n$, we can always increase its value in the objective to zero by setting $P[n]=0$ without violating the power constraints \eqref{EquPowerCon}.

Although problem (P2) resolves the non-smooth issue, it is still non-convex and difficult to solve. Based on the block coordinate descent method, we propose an iterative algorithm to solve (P2). Specifically, the optimization variables are partitioned into two blocks., i.e., $\mathbf{P}$ for transmit power control and $(\mathbf{x}, \mathbf{y})$ for UAV trajectory design. As a result, problem (P2) are divided into two subproblems: one (denoted by subproblem 1) optimizes the transmit power $\mathbf{P}$ under given trajectory $(\mathbf{x}, \mathbf{y})$; while the other (denoted by subproblem 2) optimizes the trajectory $(\mathbf{x}, \mathbf{y})$ under given transmit power $\mathbf{P}$. These two subproblems are solved in an alternating manner iteratively until the algorithm converges. The details of the proposed algorithm are given as follow.

\subsection{Subproblem 1: Optimizing Transmit Power For Given Trajectory}
Let
\begin{equation}  \label{Equa}
a_n = \frac{ \gamma_0 } {  x^2[n] + y^2[n] + H^2 },
\end{equation}
\begin{equation}   \label{Equb}
b_n = \frac{ \gamma_0  } { (x[n]-L)^2 + y^2[n] + H^2 }.
\end{equation}
Subproblem 1 can be accordingly expressed as
\begin{align}
\max_{\mathbf{P}} \; &  \sum_{n=1}^N  \ln \left( 1+ a_n P[n]   \right)  - \ln \left(1+ b_n P[n] \right) \label{EquSubProb1} \\
\text{s.t.} \; \; & \eqref{EquPowerCon} , \nonumber
\end{align}
where the property $\log_2 x = \ln x / \ln 2$ is used. Although \eqref{EquSubProb1} is a non-convex optimization problem, it has been shown in \cite{Gopala2008} that the optimal solution can be obtained as
\begin{equation}   \label{EquOptPowerScheme}
P^{*}[n] = \begin{cases}
\min \left( [ \hat{P}[n] ]^+ , P_{\text{peak}} \right)  & a_n > b_n , \\
0  & a_n \leq b_n,
\end{cases}
\end{equation}
where
\begin{equation}
\hat{P}[n] = \sqrt{ \left( \frac{1}{2b_n} - \frac{1}{2a_n}  \right)^2 + \frac{1}{\lambda } \left( \frac{1}{b_n} - \frac{1}{a_n}  \right)  }  -\frac{1}{2b_n} -  \frac{1}{2a_n}.
\end{equation}
In the above, $\lambda \geq 0$ is a constant that ensures the average power constraint $\frac{1}{N}  \sum_{ n =1 }^N P[n] \leq \bar{P}$ to be satisfied when the optimal solution of problem \eqref{EquSubProb1} is attained, which can be found efficiently via bisection search \cite{Boyd2004}.

\subsection{Subproblem 2: Optimizing Trajectory For Given Transmit Power}  \label{SecSubProb2}
Let $P_n =  \gamma_0 P[n]$. Subproblem 2 can be equivalently expressed as
\begin{align}
\max_{ \mathbf{x}, \mathbf{y} } \; &  \sum_{n=1}^N \bigg[ \ln \left( 1+ \frac{ P_n } {  x^2[n] + y^2[n] + H^2 }  \right)  \nonumber  \\
& -  \ln \left(1+ \frac{ P_n } { (x[n]-L)^2 + y^2[n] + H^2 } \right)  \bigg]  \label{EquSubProb2Ori}  \\
\text{s.t.}  \; \; & \eqref{EquMobilityCon}. \nonumber
\end{align}
Note that the objective function of problem \eqref{EquSubProb2Ori} is non-concave with respect to $\mathbf{x}$ and $\mathbf{y}$, so it cannot be solved efficiently. By introducing slack variables $\mathbf{t} = \left[ t[1], \ldots, t[N]  \right]^{\dagger}$ and $\mathbf{u} = \left[ u[1], \ldots, u[N]  \right]^{\dagger}$, we formulate the following problem
\begin{subequations}   \label{EquSubProb2Reform}
\begin{align}
\max_{ \mathbf{x}, \mathbf{y}, \mathbf{t}, \mathbf{u} } \; &  \sum_{n=1}^N \bigg[  \ln \left( 1+ \frac{ P_n } { u[n]  }  \right)  -  \ln \left(1+ \frac{ P_n } { t[n] } \right)   \bigg]\label{EquObjSub2}    \\
\text{s.t.}  \; \; \;  &  t[n] - x^2[n] + 2Lx[n] -L^2 - y^2[n] - H^2 \leq 0 , \; \forall n  \label{EquCont}  \\
&  t[n] \geq 0, \; \forall n , \label{EquContGeq0} \\
&  x^2[n] + y^2[n] + H^2 -u[n] \leq 0, \; \; \forall n,  \label{EquConu} \\
& \eqref{EquMobilityCon}.  \nonumber
\end{align}
\end{subequations}
It can be verified that with the optimal solution of problem \eqref{EquSubProb2Reform}, constraints \eqref{EquCont} and \eqref{EquConu} should hold with equalities since otherwise, $t[n]$ ($u[n]$) can be increased (decreased) to improve the objective value. Therefore, problems \eqref{EquSubProb2Ori} and \eqref{EquSubProb2Reform} have the same optimal solution of $(\mathbf{x},\mathbf{y})$. Next, we focus on solving problem \eqref{EquSubProb2Reform}.

The term $ \ln \left( 1+ \frac{ P_n } { u[n]  }  \right)$ in \eqref{EquObjSub2} and the terms $-x^2[n]$ and $-y^2[n]$ in \eqref{EquCont} make problem \eqref{EquSubProb2Reform} non-convex and difficult to be solved optimally. Based on the successive convex optimization method, we propose the following algorithm for problem \eqref{EquSubProb2Reform} to obtain an approximate solution. The algorithm assumes a given initial point $\mathbf{x}_{\text{fea}} \triangleq \left[ x_{\text{fea}}[1], \ldots, x_{\text{fea}}[N] \right]^{\dagger}$, $\mathbf{y}_{\text{fea}} \triangleq \left[ y_{\text{fea}}[1], \ldots, y_{\text{fea}}[N] \right]^{\dagger}$ and $\mathbf{u}_{\text{fea}} \triangleq \left[ u_{\text{fea}}[1], \ldots, u_{\text{fea}}[N] \right]^{\dagger}$, which is feasible to \eqref{EquSubProb2Reform}.

Since $ \ln \left( 1+ \frac{ P_n } { u[n]  }  \right)$ is a convex function of $u[n]$, its first-order Taylor expansion at $u_{\text{fea}}[n]$ is a global under-estimator, i.e.,
\begin{equation}  \label{EquTayloru}
\ln \left( 1+ \frac{ P_n } { u[n]  }  \right) \geq  \ln \left( 1+ \frac{ P_n } { u_{\text{fea}}[n]  }  \right) - \frac{ P_n ( u[n] - u_{\text{fea}}[n] ) }{  u_{\text{fea}}^2[n] + P_n u_{\text{fea}}[n]  }.
\end{equation}
Since $-x^2[n]$ and $-y^2[n]$ are concave functions, their first-order Taylor expansions at $x_{\text{fea}}[n]$ and $y_{\text{fea}}[n]$ are global over-estimators, i.e.,
\begin{equation}   \label{EquTaylorx}
-x^2[n] \leq x_{\text{fea}}^2[n] - 2 x_{\text{fea}}[n] x[n],
\end{equation}
\begin{equation}   \label{EquTaylory}
-y^2[n] \leq y_{\text{fea}}^2[n] - 2 y_{\text{fea}}[n] y[n].
\end{equation}
With \eqref{EquTayloru}--\eqref{EquTaylory}, problem \eqref{EquSubProb2Reform} is recast as
\begin{subequations}  \label{EquSubProb2Traject}
\begin{align}
\max_{ \mathbf{x},\mathbf{y}, \mathbf{t},\mathbf{u} } & \sum_{n=1}^N \bigg[ - \frac{ P_n u[n] }{   u_{\text{fea}}^2[n] + P_n u_{\text{fea}}[n]  }  -  \ln \left(1+ \frac{ P_n } { t[n] } \right)  \bigg]  \\
\text{s.t.} \; \;  & t[n] + x_{\text{fea}}^2[n]  - 2 x_{\text{fea}}[n] x[n] + 2Lx[n]  - L^2 \nonumber \\
& + y_{\text{fea}}^2[n] - 2 y_{\text{fea}}[n] y[n]   - H^2  \leq 0  , \; \forall n, \\
& \eqref{EquMobilityCon}, \; \eqref{EquContGeq0}, \; \eqref{EquConu}. \nonumber
\end{align}
\end{subequations}
Since the objective function of problem \eqref{EquSubProb2Traject} is concave, and the feasible region is convex, it is a convex optimization problem, which can be efficiently solved by the interior-point method \cite{Boyd2004}. Since the first constraint of problem \eqref{EquSubProb2Traject} implies that of problem \eqref{EquSubProb2Reform}, the solution of problem \eqref{EquSubProb2Traject} is guaranteed to be a feasible solution of problem \eqref{EquSubProb2Reform}. Moreover, since problem \eqref{EquSubProb2Traject} maximizes the lower bound of the objective function of problem \eqref{EquSubProb2Reform}, and the lower bound and the objective function of \eqref{EquSubProb2Reform} are equal only at the given point $(\mathbf{x}_{\text{fea}}, \mathbf{y}_{\text{fea}}, \mathbf{u}_{\text{fea}})$, the objective value of problem \eqref{EquSubProb2Reform} with the solution obtained by solving problem \eqref{EquSubProb2Traject} is no smaller than that with the given point $(\mathbf{x}_{\text{fea}}, \mathbf{y}_{\text{fea}}, \mathbf{u}_{\text{fea}})$.

\subsection{Overall Algorithm}  \label{SecOverallAlg}
In summary, the proposed algorithm solves the two subproblems \eqref{EquSubProb1} and \eqref{EquSubProb2Traject} alternately in an iterative manner. Since the objective value of (P2) with the solutions obtained by solving the subproblems is non-decreasing over iteration, and the optimal value of (P2) is finite, the solution by the proposed algorithm is guaranteed to converge. The details of the proposed algorithm are summarized in Algorithm 1.
\begin{algorithm}
	\caption{Proposed Algorithm for Problem (P1).}
	\begin{algorithmic}[1]
		\STATE \textbf{Initialization:} Set an initial feasible solution $( \mathbf{P}^{(0)} , \mathbf{x}^{(0)}, \mathbf{y}^{(0)} )$, an initial slack variable $\mathbf{u}^{(0)}$ and $k=0$.
		\REPEAT
		\STATE Set $k \gets k+1$;
		\STATE With given $\mathbf{P}^{(k-1)}$, set the feasible points $\mathbf{x}_{\text{fea}} = \mathbf{x}^{(k-1)}$, $\mathbf{y}_{\text{fea}} = \mathbf{y}^{(k-1)}$ and $\mathbf{u}_{\text{fea}} = \mathbf{u}^{(k-1)}$, then update the trajectory variable $(\mathbf{x}^{(k)}, \mathbf{y}^{(k)})$ and the slack variable $\mathbf{u}^{(k)}$ by solving problem \eqref{EquSubProb2Traject};
		\STATE With given $(\mathbf{x}^{(k)}, \mathbf{y}^{(k)} )$, update the transmit power control variable $\mathbf{P}^{(k)}$ by using \eqref{EquOptPowerScheme}.
		\UNTIL {The fractional increase of the objective value is below a small threshold $\epsilon$.}
	\end{algorithmic}
\end{algorithm}

\section{Numerical Results}
In this section, we provide numerical results to verify the performance of our proposed joint trajectory optimization and power control algorithm (denoted as ``TO w/ PC''), as compared to the following two benchmark algorithms: trajectory optimization without power control ``TO w/o PC'' and line-segment trajectory with optimal power control ``line w/ PC''. In the ``TO w/o PC'' algorithm, transmit power is equally allocated to time slots as $P[n]=\bar{P}, \; \forall n$, and the UAV trajectory is optimized by solving problem \eqref{EquSubProb2Traject} iteratively until convergence. In the ``line w/ PC'' algorithm, Alice flies to the point above Bob at the maximum speed, then remains at that point, and finally flies at the maximum speed in order to reach its final location by the end of the last time slot. If Alice does not have enough time to reach Bob, it will turn at a certain midway point and then fly to the final location at the maximum speed. Thus, the trajectory by this algorithm constitutes only line segments. Given this trajectory, optimal power control is implemented based on \eqref{EquOptPowerScheme}. 

The distance between Bob and Eve is set as $L=200$m, and the flying altitude is set to $H=100$m. The coordinates of Alice's initial and final locations are set as $(x_0,y_0)=(100,200)$m and $(x_F,y_F)=(100,-200)$m. The communication bandwidth is $20$MHz with the carrier frequency at $5$GHz, and the noise power spectrum density is $-169$dBm/Hz. Thus, the reference SNR at the reference distance $d_0=1$m is $\gamma_0=80$dB. The maximum speed of Alice is $v_{\max}=2$m/s. The flight time $T$ is divided into multiple time slots with equal length, and the length of each time slot is set as $d_t=0.5$s. The average power limit is set to $\bar{P}=-5$dBm, and the peak transmit power limit is $P_{\text{peak}} =4 \bar{P}$. The threshold $\epsilon$ in Algorithm 1 is set as $10^{-4}$.

\begin{figure}[!t]
	\centering
	\includegraphics[width=\columnwidth]{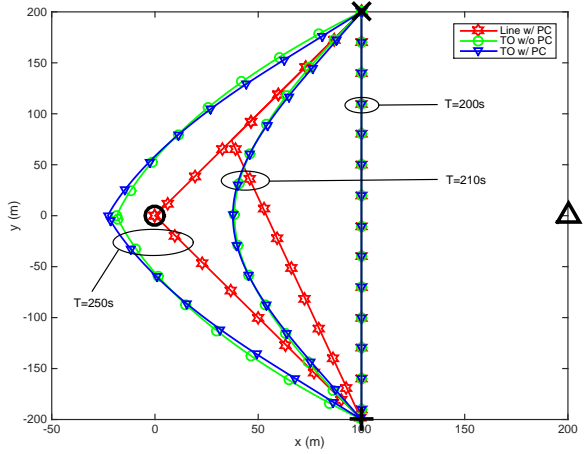}	
	\caption{Trajectories of Alice using different algorithms. }
	\label{FigTra-Ver}	
\end{figure}

\begin{figure}[!t]
	\centering
	\includegraphics[width=\columnwidth]{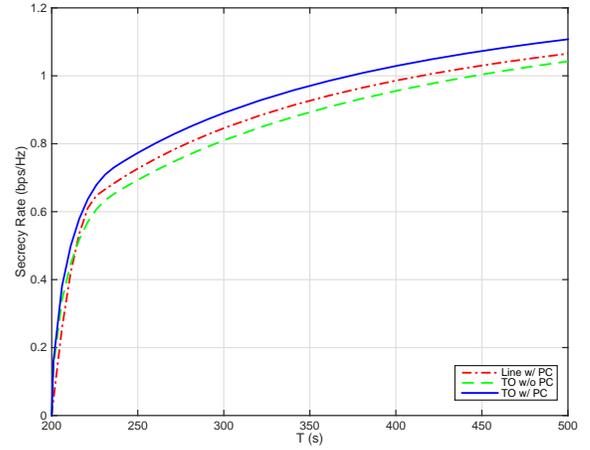}	
	\caption{Secrecy rate versus flight time $T$. }
	\label{FigSR-T-Ver}	
\end{figure}

Fig. \ref{FigTra-Ver} shows the trajectories of UAV/Alice using different algorithms with different values of $T$. The locations of Bob, Eve, Alice's initial and final locations are marked with $\bigcirc$, $\triangle$, $\times$, and $+$, respectively. It is observed that when $T=200$s, it is just enough for Alice to fly from the initial location to the final location in a straight line at the maximum speed, so the trajectories of the ``TO w/ PC'', ``TO w/o PC'' and ``line w/ PC'' algorithms are identical. The trajectories of the ``TO w/ PC'' and ``TO w/o PC'' algorithms are similar for different values of $T$, i.e., Alice tries to fly as close as possible to Bob and at the same time as away as possible to Eve as $T$ increases. When $T$ is sufficiently large, i.e. $T = 250$s, Alice first flies at the maximum speed to reach a certain location (not directly above Bob), then remains stationary at this location as long as possible, and finally flies to the final location at the maximum speed and reach there by the end of the last time slot. The hovering locations can be shown to be the best locations for maximizing the secrecy rates for the two algorithms with and without power control, respectively.

\begin{figure}[!t]
	\centering	
	\includegraphics[width=\columnwidth]{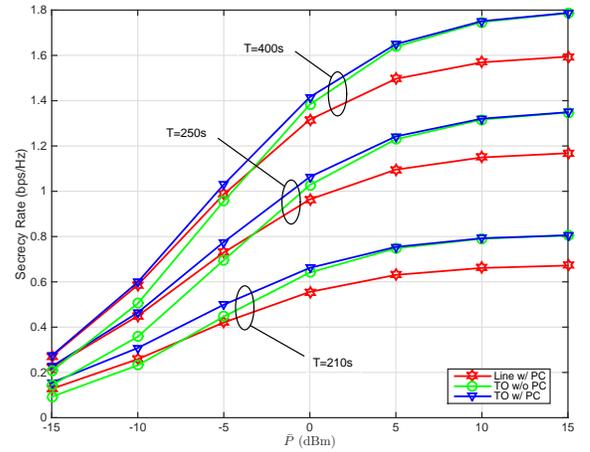}	
	\caption{Secrecy rate versus average transmit power $\bar{P}$.}
	\label{FigSR-P-Ver}	
\end{figure}

Fig. \ref{FigSR-T-Ver} shows the average secrecy rates of different algorithms versus flight time $T$. It is observed that the secrecy rates of all algorithms increase with $T$. This is because for all algorithms the maximum secrecy rate is achieved at the respective hovering locations (see Fig. \ref{FigTra-Ver}), and larger $T$ results in longer hovering time. The proposed ``TO w/ PC'' algorithm always achieves the highest secrecy rate, compared to the other two benchmark algorithms. 

Fig. \ref{FigSR-P-Ver} shows the average secrecy rates of different algorithms versus the average transmit power $\bar{P}$ with different values of $T$. It is observed that the proposed ``TO w/ PC'' algorithm always achieves the best secrecy rate. The ``line w/ PC'' algorithm achieves a secrecy rate performance close to the ``TO w/ PC'' algorithm, and significantly outperforms the ``TO w/o PC'' algorithm when $\bar{P} \leq -5$dB. This is because power control is more effective in improving secrecy rate than trajectory optimization when the average transmit power is low. It is observed that when $\bar{P}$ increases, the secrecy rate of the ``TO w/o PC'' algorithm exceeds that of the ``line w/ PC'' algorithm, and gets closer to that of the ``TO w/ PC'' algorithm. The rate gap between the ``line w/ PC'' and ``TO w/o PC'' algorithms becomes larger with increasing $\bar{P}$. This is because trajectory optimization is more effective in improving secrecy rate than power control when the average transmit power is sufficiently large. The above results demonstrate the importance and necessity of the joint trajectory optimization and transmit power control in maximizing the secrecy rate.

\section{Conclusion}
In this paper, we study the physical-layer secrecy communication in a new UAV-to-ground communication setup by exploiting the UAV trajectory design in addition to the conventional power/rate adaptation. The transmit power control and UAV trajectory are jointly designed to maximize the average secrecy rate over a finite horizon, subject to the average and peak transmit power constraints as well as practical UAV's mobility constraints. By applying the block coordinate descent and successive convex optimization methods, we propose an efficient iterative algorithm to solve this design problem. Numerical results show that the new approach of proactively controlling channel gains by adjusting the UAV trajectory with joint power control can significantly improve the physical-layer security performance of UAV communication systems.

\end{document}